# Approximation of Bandwidth for the Interactive Operation in Video on Demand System

Soumen Kanrar  
Department of Computer Science  
Vidyasagar University  
Midnapore-02, India  
rscs.soumen@mail.vidyasagar.ac.in

Niranjan Kumar Mandal  
Department of Electrical Engineering  
UEM University  
Calcutta 160, India  
niranjanmandal@uem.edu.in

*Abstract*— An interactive session of video-on-demand (VOD) streaming procedure deserves smooth data transportation for the viewer, irrespective of their geographic location. To access the required video, bandwidth management during the video objects transportation at any interactive session is a mandatory prerequisite. It has been observed in the domain likes movie on demand, electronic encyclopedia, interactive games, and educational resources. The required data is imported from the distributed storage servers through the high speed backbone network. This paper presents the viewer driven session based multi-user model with respect to the overlay mesh network. In virtue of reality, the direct implication of this work elaborately shows the required bandwidth is a causal part in the video on demand system. The analytic model of session based single viewer bandwidth requirement model presents the bandwidth requirement for any interactive session like, pause, move slow, rewind, skip some number of frames, or move fast with some constant number of frames. This work presents the bandwidth requirement model for any interactive session that brings the trade-off in data-transportation and storage costs for different system resources and also for the various system configurations.

I. INTRODUCTION

The content placement problem addressed towards the other different optimization objectives likes, Almeida et al. [1], [7], [12],[13] that directed towards the minimum hops count to minimizing the total delivery cost in the network. Zhou et al., [2] have shown the target jointly maximizing the average encoding bit rate during content transfer over the network. The overall average numbers of content replicas that are minimizing the communication load, imbalance the storage servers [7]. The brief content placement for Peer-to-Peer (P2P) Video on demand systems is presented by Suh et al.[3]. Tewari and Kleinrock [4], [5] have designed a simple queuing based model to shape the number of replicas in corresponding to the request rate of the content driven. LRU (least recent used) and LRFU (Least recent frequently used) based worked efficiently classified the batches of content with respect to the hit count. Hit Count efficiently updating the peer cache and proxy server cache in the overlay mesh type network for content replicas [10], [11]. The viewer initiative huge traffic of request to the video on demand network generally follows the Zipf like distribution. The traffic load and shape of the load inside the on demand system is presented in the articles [8], [9].Victor o.k. et.al [6] proposed different types of batching request model. The model proposed by Victor [6] integrates both the viewer activity, and the batching model. In this model the viewer requests are batching and the effect of such batching is captured in a continued batching model. The viewer activity includes the various interactive modes like pause, move slowly, move fast, etc. But the proposed model by Victor [6] is unable to describe the viewer initiated interactive operational mode. The video on demand system required a standard uniform model that can be implemented to determine the requirements of bandwidth during the viewer initiated interactive session [14]. In reality, the proper bandwidth requirement model during the interactive session obviously carries the natural trade-off in communication and storage costs for different system resource with various configurations. The paper is organized with a brief introduction at the session one, with some of the challenging issues. The basic analytic structure is described in the section two. The session based multi-user model is presented at the session three. The corresponding single viewer bandwidth requirement model is described in the session four. The session five presents the concluding remarks with the reference at the end.

II. ANALYTIC STRUCTURE

The bandwidth requirement varies during the interactive session. Viewers are using various mode of operation such as fast forward, skip, pause, and rewind. The bandwidth requirement is continuously varied with the mode of viewer driven operation. In general, let $p$ is the initial (prior) probability to select a channel according to the Binomial distribution, the total available bandwidth $B$ (i.e. in the trunk can provide) and channel required bandwidth is $W$, $(p \approx W/B)$. In reality $W$ (assuming) varies according to the viewer mode of operation within the interval $(W_{min}, W_{max})$ such that $(W_{min} \leq ... \leq W_{min+t(i)} \leq ... \leq W_{max})$ for the $i^{th}$ type of real interactive session. Here $t(i)$ is the bandwidth demand adjustment function. Initially $W_{max}$ is assigned to each channel, but in real time when huge request is submitted at a session, the channel is assigned with adjusted bandwidth or otherwise at least $W_{min}$ from the available residue. If it is impossible, then the request is buffered to the queue (at the 1st queue slot of the proxy server otherwise to any intermediate router). For connection, setup $W_{min}$ is the mandatory requirement. The bandwidth is assigned to channel according to the viewer requirement and the availability of bandwidth. Viewer requirement or the required bandwidth is maximized according to the condition, availability of bandwidth from the (trunk). During the phase of connection startup to connection closed, the viewer behavior goes through a number of various interactive sessions in a random order. The different mode of operations are denoted with notations like fast forward as ($I^1$), play normal as ($I^2$), play slow or move slowly as ($I^3$), rewind as ($I^4$), pause as ($I^5$) etc. In general, it gives $I^i \geq I^j$ for $i \geq j$ according to the bandwidth requirement. Now, there is possibility of more than one mutually exclusive outcome for every type of interactive operation during any session. Let, $e_i$ for $1 \leq i \leq k$ are $k$ mutually exclusive and exhaustive out come with the respective probabilities $p_i$ according to the viewer mode of operation. For example, $p_1$ presents the fast forward, $p_2$ presents the play normal, $p_3$ presents play slow, $p_4$ presents the rewind etc. If $e_1$ occurs $x_1$ times, $e_2$ occurs $x_2$ times $\cdots\cdots e_k$ occurs $x_k$ times assuming that $e_i$ occurs $x_i$ times for $(1 \leq i \leq k)$. Now, $m$ are the numbers of independent observations of any customer behavior during a session, then $\sum x_i = m$.

So for the $T^{th}$ customer we get,
$$P(X = X_T) = p(x_1, x_2, ....., x_k) \quad (1)$$
$$= C p_1^{x_1} p_2^{x_2} ....... p_k^{x_k}$$

Here, $C$ are the numbers of permutation of the events.

Clearly, $C = \dfrac{\lfloor m}{\lfloor x_1 \lfloor x_2 \lfloor x_3 .......... \lfloor x_k}$

Hence,

$p(x_1, x_2, ...., x_k) = \dfrac{\lfloor m}{\lfloor x_1 \lfloor x_2 \lfloor x_3 .......... \lfloor x_k} p_1^{x_1} p_2^{x_2} ....... p_k^{x_k}$

for, $0 \leq x_i \leq m$ and $\sum x_i \leq m$, it brings

$$= \dfrac{\lfloor m}{\prod_{i=1}^{K} \lfloor x_i} \prod_{i=1}^{K} p_i^{x_i} \quad (2)$$

This is true for "single customer" here the allocated bandwidth for that channel is $(\geq W_{max})$. So, $n \approx \dfrac{B}{W_{max}}$, $B$ is the total bandwidth of the trunk in ideal situation or in ideal scenario. Here, $n$ are the numbers of active links for the corresponding viewers.

III. SESSION BASED MULTI VIEWER MODEL

So, according to the function (1) the distribution is expressed as
$$P(X_1 \cap X_2 ... \cap X_n) = \prod_{T=1}^{n} P(X = X_T) \quad (3)$$

Now we proceed to calculate $p_i$ for an interactive session. Different amount of bandwidth are required for different services mode likes, pause, fast forward, rewind, play normal or play slow. The number of frame download by the viewer node varies, but the frame length is fixed for different interactive session. When the channel is assigned to the viewer then at least $(y)$ amount of bandwidth is assigned to the viewer for any interactive session here $\Rightarrow W_{min} \leq y \leq W_{max}$. It generally follows the probability distribution $p_i = \dfrac{y - W_{min}}{W_{max} - W_{min}}$ such that $\sum p_i = 1$, $p_i \geq 0$ for $\forall i \geq 0$

In any session, it follows the mixed type distribution function according to (1), (2) and (3)

$P(X_1 \cap X_2 ... \cap X_n)$

$= \prod_{t=1}^{n} \dfrac{\lfloor m}{\prod_{i=1}^{K} \lfloor x_i} \prod_{i=1}^{K} \left[ \int_{W_{min}}^{W_{max}} \left[ \dfrac{y - W_{min}}{W_{max} - W_{min}} \right]^{x_i} dy \right]$

Clearly, $n$ is the total number of viewer actively presents in the video on demand system and $m$ is the total number of interactive mode used by any customer in that session. $k$ is the any particular type of mode for example any one likes, pause, move forward, rewind etc. Here, $x_i$ are the numbers of times that modes are used by that particular viewer during that session.

IÇ. SINGLE VIEWER BANDWIDTH REQUIREMENT MODEL

Now, we present the derivation of $p_i$ that defined in section 2. Here, $i$ be any type of interactive mode of operation that is initiated by any individual viewer. The modes of interactive operations are pause, rewind, fast forward, move slow, etc. In any session, if $b_i$ is the required bandwidth for any particular interactive mode of operation then, $b_i \hbar 0$, $i \in I^{\hbar 0}$. It is common possibility that in a session, there are possible $(L \geq 1)$ numbers of interactive executable mode exist. So in a session the bandwidth requirement for a viewer mode of operation satisfy the expression (4)

$$W_{\min} \leq Max\{b_i\}_{i \in \{1,...,L\}} \approx W \quad (4)$$

During the interactive mode of operation at any session, there exist multiple sub sessions in that particular interactive session. So the occurrence of that particular interactive sub session is expressed by the following expression (5).

$$P(X_i = b_i / X_1 = b_1, ....., X_{i-1} = b_{i-1}) \quad (5)$$

To derive the expression (5), we have considered the volume of the solid sphere in $k$ dimension plane such that

$$P\left[\frac{(X_i = b_i)}{(X_1 = b_1, ....., X_{i-1} = b_{i-1}, X_{i+1} = b_{i+1}, ..., X_k = b_k)}\right] \quad (6)$$

Here, $b_i$ is the required bandwidth to continue and maintain the smooth streaming for $i^{th}$ type of active interactive operation. Here, $(c \geq 0)$ is the very minimum required bandwidth for other passive interactive viewer at the same session. So the reserved bits are required for other interactive service to maintain the smooth connectivity brings $(c \approx W_{\min})$. So, the expression (6) becomes

$$P(X_i = b_i / X_1 = c, ....., X_{i-1} = c, X_{i+1} = c, ..., X_k = c) \quad (7)$$

We assume $c$ as the required reserve bits for other interactive viewer modes of operation except $i^{th}$ type of interactive operation. In general, the $k$ dimensional volume of Euclidean ball with radius $R$ in $k$ dimensional Euclidean space is

$$V_k(R) = \frac{\Pi^{k/2}}{\lfloor(k/2+1)\rfloor} R^k$$

So, for even and odd, we obtained the following expression

$$V_{2k}(R) = \frac{\Pi^k}{\lfloor k \rfloor} R^{2k} \quad \text{and} \quad V_{2k+1}(R) = \frac{2^{k+1}\Pi^k}{\lfloor 2k+1 \rfloor} R^{2k+1}, \quad k \text{ are the}$$

number of interactive mode in a session. At every session, the volume of the sphere is continuously changing due to the interactive mode of viewer operation. According to the virtue of reality, we get the below inequality

$$V_k(R) \approx \frac{\Pi^{k/2}}{\lfloor(k/2)+1\rfloor} c^{k-1}(W_{Max}) \leq \frac{\Pi^{k/2}}{\lfloor(k/2)+1\rfloor}(W_{Max})^k$$

The expression (7) becomes

$$P(X_i = b_i / X_1 = b_1, ....., X_{i-1} = b_{i-1}, X_{i+1} = b_{i+1}, ..., X_k = b_k)$$

$$\approx \left[\frac{\Pi^{k/2}}{\lfloor((k/2)+1)\rfloor} c^{k-1}(W_{Max})\right]^{-1} \int_{R^k} dx_1 ....... dx_k \quad (8)$$

Now, we concentrate to find the bandwidth requirement for $i^{th}$ type of viewer service mode of interactive operation. As,

$$\int_{R^k} dv = \prod_{i=1}^{k}\left(\int_{-\infty}^{\infty} dx_i\right), \text{ (by considering the integration by}$$

parts) and allocated bandwidth for that viewer holds true in the following expression

$$\begin{cases} x_1 + x_2 + ........ + x_k \leq W \\ W_{\min} \, p \, x_i \leq W_{\max}, 1 \leq i \leq k \end{cases} \quad (9)$$

The interactive sessions are mutually independent for each customer. To reduce the computational complexity, we consider only one type of interactive session i.e. $i^{th}$ type like any one, normal watch or moves forward, moves backward. If a viewer opens more than one window, for example, one window is paused and another window is using different service except paused i.e. using interactive viewer mode of operation. Then the service request is multiplexed to one stream and transmitted through the already allocated channel. The expression (8) becomes

$$P(X_i = b_i /(X_1 = b_1, ....., X_{i-1} = b_{i-1}, X_{i+1} = b_{i+1}, ..., X_k = b_k) \quad (10)$$

$$\approx \left[\frac{\Pi^{k/2}}{\lfloor((k/2)+1)\rfloor} c^{k-1}(W_{max})\right]^{-1} \int_{x_i=0}^{W_{\max}} ... \int_{x_{i-1}=0}^{[W-(x_1+..+x_{i-2}+x_i+..+x_k)]}$$

$$.. \int_{x_{i+1}=0}^{[W-(x_1+..+x_i+x_{i+2}+..x_k)]} . \int_{x_k=0}^{[W-(x_1+..+x_i+...+x_{k-1})]} f(x_1,....x_k) dx_k ... dx_{i+1} dx_{i-1} dx_i$$

Here, $f(x_1,....x_k)$ is a $k$ dimensional integrand function related to $k$ –dimension volume function. In special case we can consider $f(x_1,....x_k) \approx (1)^k = 1$

ç. SIMULATION ENVIRONMENT

In simulation, we have considered the mesh overlay network from where the data be pulled to the viewer during interactive session. The number of the viewer's request inside the network grows exponentially from 0 to 1000 within one minute i.e. 60 seconds. We consider the unified demand parameter as $\approx \left(\frac{k}{n}\right)$ according to equation (10) with the expression is presented at the end of the section 2. It implies that $k$ is the number of interactive mode, used in a session with $n$ number of active links for correspond viewers. So, $n$ is depended upon the size of the network at that session.

But, $n \approx \frac{B}{W_{\max}}$, it gives a set of parameters $\approx \frac{k(w_{\max})}{B}$, here $B$ is the total bandwidth of the trunk provided in ideal situation.

The figure 1 and figure 2 presents the traffic load inside the network for various stages of the network with various statuses during the viewer driven interactive session. We have been observed from the figures that the traffic load increases inside the network with the increases number of viewer. In the simulation, the viewer's executed multiple random interactive *B* operations during any session.

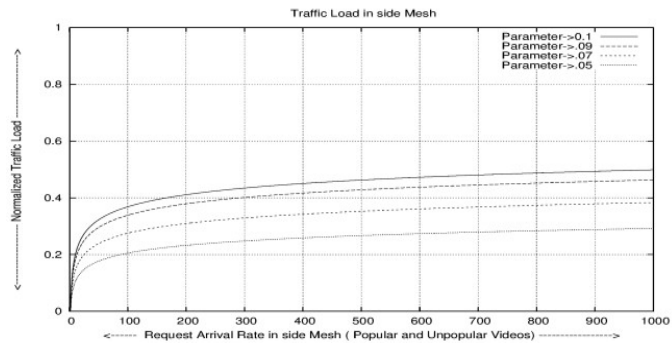

Fig. 1. Normalized traffic load in interactive session with parameter set 1

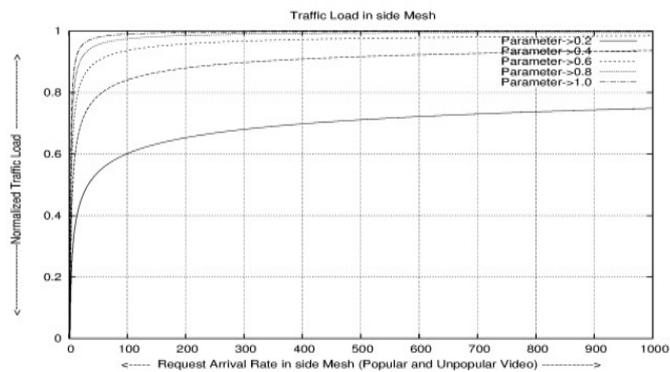

Fig. 2. Normalized traffic load in interactive session with parameter set 2

ςI.  CONCLUDING REMARKS

We have seen that the bandwidth requirement is completely depended upon the number of interactive modes used by the viewer at any session. Increase the number of viewer brings higher used of interactive modes that leads towards the high bandwidth consumption at any session. The size of the network produces little impact to this scenario. The size of the network increases linearly scalable with the number of peer's node in the network. As the unified demand parameters do not depend on the size of the network or the numbers of peer nodes. Furthermore, the single-viewer bandwidth requirement model for interactive mode at any session do not depends upon the numbers of peer node present in the mesh network. The simulation result also reflects the same scenario. As we have seen the unified demand parameter increases from 0.0 with the step size 0.2 to 1.0 in the figure 2. The normalized traffic load for the interactive session inside the overlay mesh network is presented in figure 2. A similar result is presented for the unified demand parameter value starting from 0.01 with step size 0.02 during interactive operation in the figure 1. In both the cases, we have considered the duration of the interactive session is 60 seconds.